\documentclass[preprint,proceedings]{rmaa}

\SetYear{2005}
\SetConfTitle{11th LATIN-AMERICAN REGIONAL IAU MEETING}

\title{Analysis of the Luminosity-Stellar Mass-Metallicity Relation in cosmological simulations}

\author{
  M. E. De Rossi,\altaffilmark{1,2} 
  P. B. Tissera,\altaffilmark{1,2}
  C. Scannapieco\altaffilmark{1,2}}

\altaffiltext{1}{Instituto de Astronom\'{\i}a y F\'{\i}sica del Espacio, Casilla de Correos 67, Suc. 28, 1428, Ciudad de Buenos Aires, Argentina (\protect\email{derossi@iafe.uba.ar}).}
\altaffiltext{2}{Consejo Nacional de Investigaciones Cient\'{\i}ficas y T\'ecnicas, CONICET, Ciudad de Buenos Aires, Argentina.}

\suppressfulladdresses

\listofauthors{M. E. De Rossi}
\indexauthor{De Rossi, M. E.}

\addkeyword{cosmology:theory}
\addkeyword{galaxies:formation}
\addkeyword{galaxies:evolution}
\addkeyword{galaxies:abundances}

\begin{document}
\maketitle 

\boldabstract{In this work, we study the Luminosity-Metallicity 
Relation (LMR) and the Stellar Mass-Metallicity Relation (MMR)
by using chemo-dynamical simulations in a cosmological
scenario, which allow the description of
the non-lineal growth of structure together with the chemical
enrichment of baryons.}

The simulations have been performed by employing
the chemical GADGET-2 (Scannapieco et al. 2005).
We assume a $\Lambda$CDM  cosmology
($\Omega =0.3, \Lambda =0.7, \Omega_{b} =0.04$
and $H_{0} =100 h$ km ${s}^{-1} {\rm Mpc}^{-1}, h=0.7$),
which leads to a hierarchical assembly of structure.
  
Galactic properties are estimated 
at the optical radius, defined as the one which
encloses 83 \% of the baryonic mass of the system.
Colors and magnitudes of simulated galaxies are calculated
by applying population synthesis models (see De Rossi
et al. 2006, in preparation).

Our results predict a tight correlation between
oxygen chemical abundance and luminosity for
galactic systems up to $z=3$.  Metallicity decreases
with the  absolute magnitudes of simulated galaxies
in a linear way consistently with observational trends
(e.g. Tremonti et al. 2004). Moreover, the simulated LMR
evolves with redshift in such a way that at
a given oxygen abundance systems are $\sim 3$ mag brighter
at $z=3$ than local ones, in good
agreement with observations (e.g. Kobulnicky et al. 2003).

The simulated MMR shows a similar trend to that
found by Tremonti et al. (2004) for SDSS galaxies
but with a displacement
of $\sim -0.25$ dex in the zero point.  This discrepancy 
may be associated to the fact that the SDSS 
explores only the central regions of galaxies leading
to an overestimation of their mean metal abundance.
On the other hand, at lower masses  simulated systems are
more enriched than observed galaxies which 
could probably be due to the lack of efficient
supernova energy feedback in 
our numerical model. 

The metallicity of simulated galaxies tends to
increase with stellar mass in a non-linear way.
We have determined a characteristic stellar mass 
$M_{\rm c} \sim 10^{10.2} M_{\odot} h^{-1}$ above which
the MMR gradually flattens.  This characteristic
mass seems to be independent of redshift and
segregates two galactic populations with
different astrophysical properties (Tissera et al. 2005).
It is worth mentioning that $M_{\rm c}$ agrees with the
characteristic mass for galaxy evolution previously
reported by
Kauffmann et al. (2003), Tremonti et al. (2004) and Gallazzi et al. (2005).
According to our results, systems with stellar masses
greater than $M_{\rm c}$ have suffered important merger
events and formed 
approximately
half of their present stellar mass
at $z>2$.  At lower redshifts these objects have small fractions of left 
over gas.
Low mass systems form their stars in a more passive way or
by rich gas mergers at lower $z$, leading to a stronger correlation
between stellar mass and chemical abundance.

Thus, we conclude that 
the features of the MMR may be  linked to
the hierarchical aggregation of structure in a 
$\Lambda$CDM cosmology and, hence, its study could
reveal important clues about 
how galaxies are assembled.

\end{document}